\PassOptionsToPackage{dvipsnames, table}{xcolor}
\documentclass[sigconf]{acmart}

\usepackage{color}
\usepackage[ruled,boxed,commentsnumbered, linesnumbered]{algorithm2e}
\usepackage{subcaption}
\usepackage[T1]{fontenc}
\usepackage{xspace,graphicx,fancyvrb,multirow}
\usepackage{booktabs}
\usepackage{array}
\usepackage{fancyhdr}
\usepackage{enumitem}
\usepackage[labelfont=bf,font=small, compatibility=false, skip=5pt]{caption} %
\usepackage{float}
\usepackage{syntax}

\usepackage{fp}
\usepackage{siunitx}
\usepackage{longtable}

\usepackage{listings}
\definecolor{lightgrey}{rgb}{0.92, 0.92, 0.92}
\definecolor{lightblue}{rgb}{0.4, 0.4, 0.95}

\usepackage{listings}
\DeclareCaptionType{listing}[Listing][List of Listings]

\AtBeginEnvironment{listing}{\setcounter{listing}{\value{lstlisting}}}
\AtEndEnvironment{listing}{\stepcounter{lstlisting}}

\usepackage{balance}

\usepackage[most]{tcolorbox}
\usepackage{multicol}

\newcommand{\sys}{\mbox{\textsc{MoCQ}}\xspace}
\newcommand{\numnewpattern}{46\xspace}
\newcommand{\numnewvul}{25\xspace}
\newcommand{\numPR}{5\xspace}

\newcommand{\pt}[1]{\textcolor{gray}{#1}}

\newcommand{\cc}[1]{\mbox{\smaller[0.5]\texttt{#1}}}

\fvset{fontsize=\scriptsize,xleftmargin=8pt,numbers=left,numbersep=5pt}

\input{code/fmt}

\def\Snospace~{\S{}}

\newcommand{\x}{$\times$\xspace}

\newif\ifdraft\drafttrue
\newif\ifnotes\notestrue
\ifdraft\else\notesfalse\fi

\newcommand{\eg}{{\em e.g.}}

\newcommand{\etc}{{\em etc.}}
\newcommand{\ie}{{\em i.e.}}

\input{glyphtounicode}
\newcolumntype{R}[1]{>{\raggedleft\let\newline\\\arraybackslash\hspace{0pt}}p{#1}}

\newcommand{\squishlist}{
\begin{itemize}[noitemsep,nolistsep,leftmargin=10pt]
\setlength{\itemsep}{-0pt}
}
\newcommand{\squishend}{
\end{itemize}
}

\usepackage{tikz}

\usepackage{bbding}
\usepackage{pifont}
\newcommand{\XV}{\ding{52}\textsuperscript{\kern-0.4em\ding{56}}\xspace}

\usepackage{xstring}
\newcommand{\PP}[1]{%
  \begingroup
    \edef\temp{\detokenize{#1}}%
    \IfEndWith{\temp}{.}{%
      \noindent\textbf{#1}%
    }{%
      \noindent\textbf{#1.}%
    }%
  \endgroup
}

\newcommand{\boxbeg}{
\vspace{2px}
\noindent\begin{tabular}{|l|}\hline
\begin{minipage}{3.2in}
\vspace{2px}
\noindent
}

\newcommand{\boxend}{
\vspace{2px}
\end{minipage}\\ \hline
\end{tabular}
}

\newtcolorbox{mybox}[3][float=h]
{
  colback=#2!5!white,
  colbacktitle=#2!15!white,
  boxrule=0.25mm, 
  top=0pt, bottom=0pt, left=0pt, right=0pt,
  coltitle=black,
  title    = {#3},
  #1,
}

\newcounter{prompt}
\renewcommand{\theprompt}{\arabic{prompt}}  %

\newcommand{\promptbox}[2]{ %
\refstepcounter{prompt}
\begin{mybox}{yellow}{\textbf{Prompt \theprompt: #1}}
{#2}
\end{mybox}
}

\newcommand{\answerbox}[2]{%
\begin{mybox}{yellow}{#1}
{#2}
\end{mybox}
}

\newcommand{\appautoref}[1]{\hyperref[#1]{Appendix~\ref{#1}}}

\begin{document}

\title{Neuro-symbolic Static Analysis with LLM-generated Vulnerability Patterns}
\settopmatter{printacmref=false, printfolios=true}

\copyrightyear{2025}
\acmYear{2025}

\author{Penghui Li}
\affiliation{
\institution{Columbia University}
\city{New York, NY}
\country{USA}
}
\email{pl2689@columbia.edu}

\author{Songchen Yao}
\affiliation{
\institution{Columbia University}
\city{New York, NY}
\country{USA}
}
\email{sy2743@columbia.edu}

\author{Josef Sarfati Korich}
\affiliation{
\institution{Columbia University}
\city{New York, NY}
\country{USA}
}
\email{jsk2266@columbia.edu}

\author{Changhua Luo}
\affiliation{
\institution{Wuhan University}
\city{Wuhan, Hubei}
\country{China}
}
\email{chluo2502@whu.edu.cn}

\author{Jianjia Yu}
\affiliation{
\institution{Johns Hopkins University}
\city{Baltimore, MD}
\country{USA}
}
\email{jyu122@jhu.edu}

\author{Yinzhi Cao}
\affiliation{
\institution{Johns Hopkins University}
\city{Baltimore, MD}
\country{USA}
}
\email{yinzhi.cao@jhu.edu}

\author{Junfeng Yang}
\affiliation{
\institution{Columbia University}
\city{New York, NY}
\country{USA}
}
\email{junfeng@cs.columbia.edu}

\settopmatter{authorsperrow=4}

\renewcommand\footnotetextcopyrightpermission[1]{}


\begin{abstract}
In this work, we present \sys, a neuro-symbolic static analysis framework that leverages large language models (LLMs) to automatically generate vulnerability detection patterns.
This approach combines the precision and scalability of pattern-based static analysis with the semantic understanding and automation capabilities of LLMs.
\sys extracts the domain-specific languages for expressing vulnerability patterns and employs an iterative refinement loop with trace-driven symbolic validation that provides precise feedback for pattern correction.
We evaluated \sys on 12 vulnerability types across four languages (C/C++, Java, PHP, JavaScript).
\sys achieves detection performance comparable to expert-developed patterns while requiring only hours of generation versus weeks of manual effort.
Notably, \sys uncovered \numnewpattern new vulnerability patterns that security experts 
had missed and discovered \numnewvul previously unknown vulnerabilities in real-world applications.
\sys also outperforms prior approaches with stronger analysis capabilities and broader applicability.
\end{abstract}

\date{}
\maketitle

\sloppy
\section{Introduction}
\label{s:intro}

Pattern-based static analysis has become a cornerstone of modern vulnerability detection.
Compared to dynamic techniques~\cite{aflgo, witcher, cafl}, it scalably analyzes large codebases with high code coverage and requires minimal setup effort (\eg{,} no need for software deployment).
It has been widely adopted in production systems~\cite{phpjoern, joern, recurscan, silent-spring} and has discovered numerous high-impact vulnerabilities (\eg{,} Log4Shell~\cite{log4shell, cve2021-44228}).
For instance, CodeQL~\cite{codeql} is extensively integrated into GitHub's CI/CD pipelines and is employed to analyze large-scale applications such as the Chrome browser~\cite{chromium-codeql}.
Joern~\cite{joern, phpjoern} has been recognized as one of the most widely used tools for automated vulnerability discovery~\cite{shiri2024systematic}.

However, this success comes at a cost. 
The effectiveness of static approaches relies on the quality of \emph{vulnerability patterns}.
These patterns describe vulnerable code structures or behaviors and are used to match new vulnerability instances in target codebases.
The creation of these patterns is largely manual and demands substantial human effort and expertise.
This process is not only time-consuming (\eg{,} several weeks or more~\cite{froberg2023detection}), but also prone to inaccuracies due to limited human knowledge.
This significantly reduces the effectiveness of detection~\cite{codeql-buggy1, codeql-buggy2}.
As codebases evolve and new threats emerge, keeping detection patterns accurate and up to date requires ongoing maintenance and engineering effort.

To alleviate this burden, researchers have alternatively explored learning-based approaches that analyze source code.
Most learning based methods operate at the function level to classify functions as vulnerable or benign. 
However, they often lack interprocedural or project-wide context~\cite{li2021sysevr, russell2018automated}, and require a large volume of labeled vulnerabilities for training~\cite{chakraborty2021deep}, which may not be available.
More recently, several works~\cite{zhou2024large, lu2024grace, robin2025} have used Large Language Models (LLMs) to directly analyze code without relying on traditional program analysis.
Leveraging massive training corpora, LLMs can recognize typical (vulnerable) usage patterns and reason about code semantics across diverse programming contexts.
While promising, LLMs face challenges such as token length limits, inconsistent reasoning, and susceptibility to hallucinations.

The distinct challenges of classic and learning-based methods suggest that a hybrid \emph{neuro-symbolic approach}\footnote{Neuro refers to learning-based analysis (particularly LLMs) leveraging neural models for semantic reasoning; symbolic refers to classic static analysis using symbolic representations and formal logic systems.} is a promising direction.
While recent works have taken steps toward such integration, they fall short of the automation level or analysis capability.
IRIS~\cite{iris} and Artemis~\cite{artemis} use LLMs to extract taint sources and sinks, and incorporate them into existing detection patterns for Java and PHP applications, respectively.
They automate only \emph{parts} of the detection logic, while the core vulnerability patterns must still be manually crafted by security experts.
For instance, IRIS expects experts to manually design the complete data-flow tracking logic and control-flow conditions in CodeQL, leaving only placeholder slots for sources/sinks to be filled by the LLM.
This means that for each new vulnerability type, experts must first spend days to weeks developing the base query structure before IRIS can contribute.
KNighter~\cite{knighter} is a concurrent work that synthesizes Clang static analyzers~\cite{csa} for C/C++ vulnerability detection.
However, as explicitly acknowledged in their paper, KNighter does not support temporal and interprocedural analysis.
Lacking such essential capabilities limits KNighter to simple, intraprocedural vulnerabilities.

In this paper, we design \sys (\underline{Mo}del-Generated \underline{C}ode \underline{Q}ueries), a general neuro-symbolic static vulnerability detection framework that addresses these limitations.
\sys leverages LLMs to automatically generate \emph{complete end-to-end vulnerability detection patterns} from scratch, including sources, sinks, vulnerable operations, control-flow logic, and sanitization checks, rather than just taint specifications.
It incorporates these LLM-generated patterns into established static analysis platforms (Joern and CodeQL), preserving their scalability and precision while automating what was previously labor-intensive.
Unlike prior approaches, \sys supports multiple programming languages and handles complex analysis tasks within a unified framework.

However, such a pattern generation solution faces multiple challenges.
LLMs often do not have enough knowledge of domain-specific languages (DSLs) used by static vulnerability detection tools (\eg{,} Scala-like DSL for Joern~\cite{joern} and SQL-like DSL for CodeQL), resulting in \emph{uncompilable queries}.
Specifically, our study shows that pretrained, commercial LLMs, \eg{,} GPT-4o~\cite{gpt4o}, often fail to generate valid queries, frequently producing \emph{syntax errors} or triggering \emph{execution-time exceptions}.
To deal with this challenge, \sys adopts a technique, specifically \emph{DSL subsetting}, to refine DSL by selecting a core set of features to guide the automated query generation, without sacrificing expressiveness.
The intuition is that such a refined subset can be better understood by pretrained LLMs while still preserving the core semantics due to the redundant nature of the DSL.

Even when the LLM-generated queries are syntactically correct and executable, they may fail to capture vulnerabilities effectively.
Specifically, they may trigger semantic errors (\ie{,} fail to report vulnerabilities), be overly specific (\ie{,} overfitting to known instances via matching exact variable names or fixed control flow paths), or be overly general (\ie{,} underfitting by matching many irrelevant code cases).
To deal with these, \sys adopts a \emph{feedback-driven} approach, which incorporates fine-grained feedback from a symbolic query validator for iterative query refinement.
The query validator executes the LLM-generated queries and monitors the exhibited behaviors, including syntax errors, execution exceptions, and semantic errors.
Specifically, we instrument the query execution runtime to obtain a block-level program state of the query execution to precisely locate the errors and inconsistencies.
These are then used for the LLM to refine the incorrect query.
The validator further uses heuristics to detect if a query is too specific and instructs the LLM to generalize it;
it also instructs LLMs to eliminate false positives (if any) to improve the precision.

We implemented \sys for two state-of-the-art static analysis tools, Joern~\cite{joern} and CodeQL~\cite{codeql}.

We extensively evaluated \sys on 12 representative vulnerability types (\eg{,} use-after-free, deserialization, prototype pollution) across four popular programming languages (C/C++, Java, PHP, and JavaScript).
\sys could efficiently generate detection queries within hours.
\sys-generated queries demonstrate strong detection capability comparable to state-of-the-art approaches and successfully detected \numnewvul new vulnerabilities in real-world applications.

Through manual analysis to compare the patterns from \sys and experts, we found that \sys uncovered \numnewpattern new vulnerability patterns missed by experts, with \numPR patterns already integrated into upstream tool repositories. 
Each missed pattern can result in overlooking a class of vulnerabilities.
Our further analysis revealed complementary strengths between LLMs and human experts.
LLMs excel at capturing subtle linguistic nuances that experts consistently overlook, while experts demonstrate superior high-level reasoning for complex vulnerabilities.
We thus combined both approaches and demonstrated that the integration could achieve better performance than either approach alone.

This paper makes the following contributions.

\squishlist
\item We propose a neuro-symbolic framework that automates end-to-end vulnerability pattern generation by leveraging DSL subsetting and feedback-driven refinement.

\item We develop \sys for Joern and CodeQL, demonstrating automated vulnerability pattern generation across multiple languages.
\item 
 We demonstrate \sys achieves detection performance comparable to state-of-the-art approaches.
 \sys discovered \numnewpattern new patterns and \numnewvul new vulnerabilities.

\squishend

\section{Background and Motivation}
\label{s:background}

\subsection{Pattern-based Static Analysis}
Pattern-based static analysis is a representative and practical approach among various static analysis techniques.
Tools (\eg{,} Joern~\cite{joern, phpjoern} and CodeQL~\cite{codeql}) express vulnerability patterns as \emph{structured queries} written in tools' DSLs.
They use language-specific frontends to translate source code into graph-like program representations such as control-flow graphs and code property graphs (CPG)~\cite{joern}.
Security analysts then write precise queries that operate on these representations to identify vulnerabilities.
Joern models programs with CPGs~\cite{joern-code}, while CodeQL relies on code databases.
The DSLs differ as well.
Joern employs a Scala-like language, whereas CodeQL uses SQL-like syntax.
Static analyzers enforce strict syntax and semantic checks during query execution to ensure accuracy and soundness.

\PP{Example: A Prototype Pollution Query}
JavaScript is a prototype-based language where objects inherit properties and methods through a mechanism known as \emph{prototypes}.
Modifying the prototype of one object can affect all other objects that inherit from the same prototype, resulting in \emph{prototype pollution}.
For example, an attacker can access the prototype via \cc{p=obj["__proto__"]} and modify with it using \cc{p["toString"]="Hacked"}.
As a result, any object inheriting from the fundamental prototype \cc{Object.prototype} now has tampered \cc{toString} method.
This type of vulnerability can lead to arbitrary code execution, privilege escalation, and denial of service~\cite{silent-spring, objlookup, cornelissen2024ghunter, liu2024undefined, kang2022probe}.

\begin{listing}[t]
    \begin{Verbatim}[commandchars=\\\{\},codes={\catcode`\$=3\catcode`\^=7\catcode`\_=8\relax}]
\PY{c+c1}{// find obj[\PYZdq{}\PYZus{}\PYZus{}proto\PYZus{}\PYZus{}\PYZdq{}] in p=obj[\PYZdq{}\PYZus{}\PYZus{}proto\PYZus{}\PYZus{}\PYZdq{}]}
\PY{k}{def}\PY{+w}{ }\PY{n+nf}{objProto}\PY{+w}{ }\PY{o}{=}\PY{+w}{ }\PY{n}{cpg}\PY{p}{.}\PY{n}{call}
\PY{+w}{  }\PY{p}{.}\PY{n}{where}\PY{p}{(}\PY{n}{\PYZus{}}\PY{p}{.}\PY{n}{name}\PY{p}{(}\PY{n+nc}{Operators}\PY{p}{.}\PY{n}{assignment}\PY{p}{)}\PY{p}{)}\PY{p}{.}\PY{n}{argument}\PY{p}{(}\PY{l+m+mi}{2}\PY{p}{)}
\PY{+w}{  }\PY{p}{.}\PY{n}{isCall}\PY{p}{.}\PY{n}{arrayAccess}\PY{+w}{ }

\PY{c+c1}{// find p[\PYZdq{}toString\PYZdq{}] in p[\PYZdq{}toString\PYZdq{}]=\PYZdq{}Hacked\PYZdq{}}
\PY{k}{def}\PY{+w}{ }\PY{n+nf}{objProp}\PY{+w}{ }\PY{o}{=}\PY{+w}{ }\PY{n}{cpg}\PY{p}{.}\PY{n}{call}
\PY{+w}{  }\PY{p}{.}\PY{n}{where}\PY{p}{(}\PY{n}{\PYZus{}}\PY{p}{.}\PY{n}{name}\PY{p}{(}\PY{n+nc}{Operators}\PY{p}{.}\PY{n}{assignment}\PY{p}{)}\PY{p}{)}\PY{p}{.}\PY{n}{argument}\PY{p}{(}\PY{l+m+mi}{1}\PY{p}{)}
\PY{+w}{  }\PY{p}{.}\PY{n}{isCall}\PY{p}{.}\PY{n}{arrayAccess}\PY{+w}{ }

\PY{c+c1}{// find a data\PYZhy{}flow path from objProp to objProto}
\PY{n}{objProp}\PY{p}{.}\PY{n}{array}\PY{p}{.}\PY{n}{reachableBy}\PY{p}{(}\PY{n}{objProto}\PY{p}{)}\PY{p}{.}\PY{n}{filter}\PY{p}{(}\PY{p}{.}\PY{p}{.}\PY{p}{.}\PY{p}{)}\PY{+w}{ }
\end{Verbatim}

    \caption{Example real-world query in Joern, simplified for clarity.}
    \label{code:pattern}
\end{listing}

\autoref{code:pattern} shows a simplified version of a real-world Joern query designed to detect JavaScript prototype pollution~\cite{froberg2023detection}.\footnote{The original query is more complex, including checks for user-controlled values, property filtering, \etc{}}
It comprises three main steps:
(1) identifying the object prototype (lines 1-4),
(2) identifying the property access (lines 6-9), 
and (3) verifying whether there is a data connection between them (lines 11-12).
For example, step (1) uses Joern's dot operator (\cc{"."}) to chain operations, where each stage filters results produced by the previous one. 
It begins by selecting assignment expressions via \cc{Operators.assignment} (\cc{assignment} is modeled as a form of \cc{call} operation in Joern)
extracts the right-hand side via \cc{argument(2)}, and narrows results to array-style accesses using \cc{arrayAccess()}.

Writing high-quality patterns is non-trivial.
For example, the query in \autoref{code:pattern} took about three weeks to draft and an additional month to refine~\cite{query-too-long}.
Experts may miss attack variants or write overly general patterns, causing false negatives and false positives~\cite{codeql-buggy1, codeql-buggy2}.
For instance, attackers can bypass the detection by combining property access and assignment in one expression, like \cc{obj["proto"]["toString"]="Hacked"}.
This also happens in the official CodeQL queries maintained by GitHub Security.

This work thus explores an automated approach to generating vulnerability patterns using LLMs.
When a new threat is publicly disclosed, a developer can rapidly generate comprehensive vulnerability patterns to scan their codebases, without the long delays traditionally required for manual pattern engineering.

\answerbox{Incorrect LLM output:}{
\begin{Verbatim}[commandchars=\\\{\},codes={\catcode`\$=3\catcode`\^=7\catcode`\_=8\relax}]
\PY{p}{.}\PY{p}{.}\PY{p}{.}
\PY{k+kd}{val}\PY{+w}{ }\PY{n}{assign}\PY{+w}{ }\PY{o}{=}\PY{+w}{ }\PY{n}{code}\PY{p}{.}\PY{n}{call}\PY{+w}{ }\PY{c+c1}{// omitting `cpg`}
\PY{+w}{  }\PY{p}{.}\PY{n}{nameExact}\PY{p}{(}\PY{l+s}{\PYZdq{}assignment\PYZdq{}}\PY{p}{)}\PY{+w}{ }\PY{c+c1}{// using wrong operator name}
\end{Verbatim}

}

\subsection{Challenges}
\label{s:problem-challenges}
While using LLMs to generate queries may seem intuitive, achieving correct and effective query generation faces substantial challenges.
To illustrate, we show an example where GPT-4o~\cite{gpt4o} was tasked with generating a Joern query to detect JavaScript prototype pollution using a chain-of-thought prompting strategy.
Despite step-by-step reasoning guidance in the prompt, the generated query failed to execute in Joern's analysis engine.
Below, we present the prompt and the failed query, followed by a summary of four key technical challenges that \sys addresses.

\PP{DSL Syntax Correctness}
The LLM-generated queries frequently raise \emph{grammar and syntax errors}, which stem from the model's incomplete understanding of the DSL grammar.
Specifically, LLMs typically lack sufficient exposure to the syntax, semantics, and typical usage of complex DSLs.
For example, Joern queries based on the code property graph usually require the root object \cc{cpg}.
However, the query generated above did not follow this requirement.
This issue is further aggravated by the hallucinations of LLMs.

\PP{Runtime Execution Correctness}
Queries must execute within the domain-specific runtime (\ie{,} analysis engine) and interact meaningfully with structured program representations.
LLM-generated queries often use incorrect or non-existent function calls, missing parameters, and incompatible data types or operations, resulting in \emph{execution-time exceptions}.
For example, the \cc{nameExact} operation in the above query used \cc{"assignment"} instead of \cc{"<operator>.assignment"}, which led to execution-time exceptions after we first manually fixed the syntax error.

\PP{Semantic Validity}
Even when queries are syntactically correct and executable, ensuring their semantic validity remains challenging.
They must satisfy runtime constraints and capture the correct traversal logic.
Executable queries may still fail to reflect intended program behaviors, leading to \emph{semantic errors}, where the query runs without crashing but misses the vulnerable code.

\section{\sys}
\label{s:design}
\begin{figure*}[t]
    \center{\includegraphics[width=\textwidth]{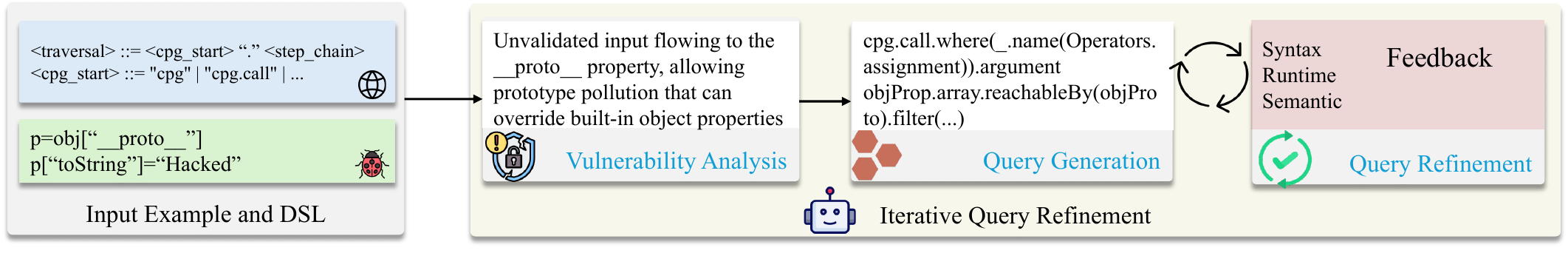}}
    \caption{The workflow of \sys.}
    \label{fig:arch}
\end{figure*}

\label{s:overview}
We design \sys, a novel neuro-symbolic static analysis system.
Its workflow is outlined in \autoref{fig:arch}.
\sys first analyzes open knowledge, such as documentation and tool implementations, to extract the query DSL (\autoref{s:design-dsl}). 
\sys then generates vulnerability queries through a feedback loop, with a trace-driven symbolic query validator to refine (\autoref{s:design-querygen}) and optimize (\autoref{s:design-opt}) the queries.
Finally, \sys outputs the generated query, which can be applied to detect new vulnerabilities.
We summarize two key techniques that solve the aforementioned challenges.

\PP{Query DSL Extraction and Subsetting}
To help generate syntactically- and semantically-correct queries, we first obtain the formal specifications of the query DSLs used by the static analysis tools, which also define the desired output structure for LLMs.
\sys extracts the DSL specifications---including DSL grammars, data types, compatible APIs, and their functionalities---by automatically parsing the online documentation and implementations.
However, directly using all extracted DSL specifications would overwhelm the LLM because of the DSL's complexity, and would still not enable the  LLM to efficiently generate valid queries.
We thus propose \emph{DSL subsetting}, which selects a core subset of DSL features from the extracted full set.
This is based on our observation that query DSL often contains redundant features, for example, those that can be equally expressed by other language features. 
Such a DSL subset could be better understood by the LLMs to significantly reduce the query generation complexity.

\PP{Iterative Feedback Loop}
Atop the core DSL subset, \sys employs an LLM to generate queries and iteratively refines them based on fine-grained feedback from our symbolic query validator.
Specifically, we instrument query execution runtime, which executes and analyzes the generated queries to locate the exact locations or causes of the syntax and semantic errors.
Beyond that, the validator also tracks the intermediate program states on the query execution traces and validates whether the generated query could retrieve the given vulnerability examples.
During such an iterative process, \sys is able to gradually generate valid queries that can analyze complex real-world programs.
Moreover, the validator incorporates heuristics to detect potential overfitting in queries and guides the LLM to generalize them.
It also instruments the LLM to refine the query to eliminate the false positives.

\subsection{DSL Specification Extraction}
\label{s:design-dsl}

To help LLMs generate syntactically and semantically valid queries, \sys extracts formal specifications of the query DSLs used by static analysis tools.
Rather than using few-shot learning (which provides a few query examples) or fine-tuning (which requires a large amount of labeled data), we guide the model using a concise DSL specification.
This specification includes both the \emph{grammar}, which defines the structural rules of the query language, and the \emph{semantics}, which describe the meaning and behavior of APIs, data types, and parameters.  
The prompts we used for DSL extraction are shown in \appautoref{s:appendix-prompts}.

\subsubsection{DSL Extraction}
\label{s:design-dsl-extraction}
To extract the DSL, \sys first analyzes the static analysis tool's documentation (\eg{,} online use instructions) and source code (\eg{,} API implementations and comments).
Documentation typically contains structured content (\eg{,} HTML tables) that describes grammar, operations, and usage patterns.
We first crawl the online documentation of the tools and prompt an LLM to extract grammar rules, data types, operations, and API usage.
For example, Joern's website~\cite{joern-doc} lists the query operations (termed ``Steps'') along with their descriptions.
Beyond the brief APIs' behaviors in the documentation, we further seek more precise definitions from their code implementations, which offer ground-truth definitions through function signatures, type annotations, inline comments, \etc{}
Based on the results of document analysis (\eg{,} keywords, API name), we prompt an LLM to retrieve relevant source code files from each tool's implementation, then to automatically summarize the functionality of the related functions.

\subsubsection{DSL Subsetting}

The extracted query DSL specification is complex and could overwhelm the model.
For example, static analysis tools like Joern and CodeQL have thousands of APIs.
Each API can have multiple parameter choices, resulting in an exponential number of combinations of these rules and operations.
Even though \sys has known the precise description for each operation, our evaluation (details in \autoref{s:eval-ablation}) demonstrates that providing LLMs with the full-set DSL does not efficiently produce good queries.

We design a \emph{DSL subsetting} technique to resolve the DSL complexity issue.
We observe that the DSL specification contains significant redundancy not strictly necessary for constructing useful queries.
A small, expressive subset is often sufficient for most vulnerability detection tasks.
Instead of exposing the entire DSL, we restrict query generation to this smaller, more tractable subset, which reduces complexity and improves accuracy.
Concretely, the subsetting process operates at the granularity of APIs, which can strike a balance between expressiveness and control---it preserves flexibility for query construction while keeping the DSL surface concise and understandable.
\sys first associates typical parameter choices with each API.
It then analyzes the full set of DSL APIs and prompts an LLM to decide and select these important, necessary DSL APIs for common usage.
This is feasible since the LLM can refer to features' intended behaviors or functionalities to evaluate if one is necessary.

\setlength{\grammarparsep}{3pt}
\begin{figure}
\footnotesize
\input{fig/grammar}
\caption{Simplified BNF grammar for Joern, extracted by \sys.}
\label{joern-grammar-simplified}
\end{figure}

\autoref{joern-grammar-simplified} presents a simplified grammar for Joern using the Backus-Naur Form (BNF)~\cite{bnf}.
It details the fundamental grammar rules, such as the \cc{cpg_start} that initializes a \cc{traversal} query, and how various components connect to form complete expressions.
The grammar rules explicitly define the structure and permissible compositions of the DSL, guiding models to produce syntactically valid statements.
For example, by following the grammar, an LLM knows to begin queries with \cc{cpg} followed by appropriate method chaining through \cc{step_chain} elements.
The grammar also clarifies how to construct complex queries using operations like \cc{filter_step} and \cc{complex_step}, enabling more sophisticated code analysis patterns.

\subsection{Query Generation and Refinement}
\label{s:design-querygen}

Atop the core DSL subset, \sys generates detection queries as described in \autoref{alg:query-gen}.
For each run, \sys is tasked with producing a query $Q^t$ that targets a specific vulnerability type $t$.  
The generator takes as input 
(1) a task description $D^t$, which provides a natural-language explanation of the vulnerability type,
and (2) a small set of representative vulnerability examples $E^t=\{e^t_i \mid i=1,\ldots,m\}$.  
\sys first constructs a program slice for each vulnerability example.
These examples both help LLMs learn the semantic patterns characteristic of a vulnerability type and serve as ground-truth test cases to validate generated queries. 
To make generation more tractable, \sys initially produces a per-example query $Q^t_i$ that is expected to retrieve the corresponding example $e^t_i$. 
The system then performs optimizations to address overfitting and underfitting (line 8). 
Finally, the system merges all per-example queries into a unified per-type query ($Q^t=\bigcup_{i=1}^{m} Q^t_i$) for effective real-world vulnerability detection (line 14).

The DSL subset is provided to \sys but omitted from the pseudocode for clarity.
Concretely, we present the core DSL subset to the LLM to guide query generation.
We adopt a \emph{grammar prompting} technique~\cite{nips-dsl} to concisely express the grammar.
Specifically, we supply an abstract grammar specification of the DSL using BNF~\cite{bnf}, which is a standard way of defining language syntax.
Although BNF is itself domain-specific, it appears more frequently in training data than custom query DSLs.
Moreover, BNF structure is simpler and more concise than real query examples, while still conveying rich information about DSL usage.
We also specify the subset semantics like APIs to the LLM, including their signature, data types, and API functionalities.

\SetAlCapNameFnt{\small}
\SetAlCapFnt{\small}
\begin{algorithm}[t]
    \caption{Iterative query generation in a feedback loop.}
    \footnotesize
    \label{alg:query-gen}
    \SetKwProg{Fn}{function}{:}{end}
    \SetKwInOut{Input}{Input}
    \SetKwInOut{Output}{Output}
    \SetKw{Break}{break}
    \SetKw{not}{not}
    \SetKwFunction{LLMQueryGen}{LLMQueryGen}
    \SetKwFunction{Validate}{Validate}
    \SetKwFunction{Execute}{Execute}
    \SetKwFunction{Optimize}{Optimize}

    \Input{Task description $D^t$, vulnerability examples $E^t$}
    \Output{Query $Q^t$}
    $QList \leftarrow []$\\
    \ForEach{$e^t_i \in E^t$}
    {
        $PState \leftarrow null$\\
        \For{$attempt \leftarrow 1$ \KwTo $MaxN$}
        {
            $Q^t_i \leftarrow$ \LLMQueryGen($e^t_i$, $D^t$, PState)\\
            error, PState $\leftarrow$ \Validate($Q^t_i$, $e^t_i$)\\
            \If{\not error}
            {
                $Q^t_i \leftarrow$ \Optimize($Q^t_i$)\\
                \texttt{QList.add}($Q^t_i$)\\
                \Break
            }
        }
    }
    $Q^t \leftarrow \bigcup_{q \in QList} q$ \tcp*[h]{\scriptsize{Merge per-example queries}}\\
    \Return $Q^t$\\

    ~\\
    \Fn{\Validate(Q, e)}
    {
        $S \leftarrow null$\\
        \ForEach{$B \in Q$}
        {
            $S \leftarrow$ \Execute($B$, $e$, $S$) \tcp*[h]{\scriptsize{Capture fine-grained runtime info.}}\\
        }
        \Return $S.err, S.State$
    }
\end{algorithm}

\subsubsection{Vulnerability Analysis}
As an initial step, \sys analyzes the input vulnerability example to understand its root cause and detection requirements.
Given a vulnerability example $e^t_i$, \sys prompts an LLM agent to systematically examine the vulnerable code and identify the key security-relevant elements.
\sys applies a chain-of-thought prompting strategy to decompose the vulnerability analysis into interpretable components.
This analysis phase extracts essential vulnerability characteristics, such as the dangerous operations, data flow paths, and missing sanitization that inform subsequent query generation.

Based on this understanding, \sys creates a structured detection plan that breaks down the overall task into smaller, well-defined subtasks.
For example, to detect prototype pollution vulnerabilities, the plan includes key steps such as: (1) identifying object property modifications, (2) tracking user-controlled data flow to property names, (3) checking for dangerous property values (\eg{,} \cc{\_\_proto\_\_}), and (4) verifying sanitization presence.
Each subtask corresponds to a specific aspect of the vulnerability pattern and can be systematically translated into DSL operations.
Importantly, \sys performs this decomposition automatically without requiring manual specification of the detection steps.
This structured plan not only makes query generation more manageable but also provides a clear framework for iterative refinement when validation feedback reveals errors or incomplete coverage.

\subsubsection{Query Refinement with Symbolic Validation}
\label{s:design-validation}
Given the difficulty of query generation, it is almost impossible to produce a qualified query in a single LLM attempt, even with the DSL.
Therefore, \sys incorporates an iterative feedback loop (shown as the \cc{for} loop on lines 4-12 of \autoref{alg:query-gen}) to debug and refine the query.
At each iteration, \sys executes the generated query and validates it across syntax compliance, execution correctness, and semantic accuracy.
When validation fails, \sys captures fine-grained feedback such as error locations, intermediate program states, and provides this information to the LLM for refinement and further fixes.
This process terminates either when a valid query is generated or when the maximum attempt threshold is reached.

To enable precise debugging, \sys represents each query $Q^t_i$ as a sequence of $n$ blocks $[B_1, B_2, \ldots, B_n]$, where each block corresponds to a detection subtask.
By instrumenting the query execution runtime, \sys records intermediate program states after executing each block $B_j$, formulated as $S_j=\{(var, val_{j}) \mid var \in B_{\leq j}\}$, capturing in-scope variables and their runtime values.
These annotated program states $PState=[S_1, S_2, \ldots, S_n]$ allow the LLM to identify exactly where and why the query fails, rather than receiving only coarse-grained boolean feedback.
We detail each validation dimension below.

\PP{Syntax Validation}
Syntax validation checks if the query complies with the extracted DSL grammar.
The static analysis tool itself provides a runtime for the queries and is naturally equipped with grammar validation capability.
However, it is often coarse-grained without pinpointing the exact error locations.
Therefore, we instrument and hook the grammar validator to obtain the corresponding grammar rule of the errors.
Alternatively, one can leverage off-the-shelf grammar parsers (\eg{,} ANTLR~\cite{parr2013definitive}).

\PP{Execution Validation}
\sys refers to the collected DSL specifications to correct execution exceptions.
There are multiple types of semantic errors, such as undefined variables, missing attributes, type mismatches, or incorrect API usage.
Our symbolic validator checks these behaviors and outputs the error locations.
Besides merely reporting the errors, \sys searches the collected DSL semantics to suggest correct ones by fuzzy matching name similarities.
For example, it could suggest \cc{"<operator>.assignment"} as a similar argument for \cc{"assignment"} to fix the \cc{exactName} exception mentioned in \autoref{s:problem-challenges}.

\PP{Semantic Validation}
Inspired by dynamic software testing~\cite{aflgo}, \sys employs trace-driven validation to provide fine-grained semantic feedback to the LLM.
\sys evaluates the semantic behaviors of query execution by debugging its runtime behavior against the vulnerability example $e^t_i$.
A straightforward approach is to check whether the query as a whole retrieves the vulnerability example (\ie{,} boolean feedback).
However, this is insufficient for helping an LLM pinpoint the root cause when the query fails to retrieve the example.
\sys tracks intermediate program states along the query execution trace to enable fine-grained debugging.
For example, in the prototype pollution query in \autoref{code:pattern}, after executing the block that identifies \cc{obj["\_\_proto\_\_"]} assignments (variable \cc{objProto}), \sys captures concrete code locations as a tuple of (\cc{objProto}, \cc{[Node@line4]}).
If \cc{objProto} is empty when it should detect the vulnerability example, this signals that the pattern is incorrectly formulated.
\sys provides the LLM with the annotated trace values $PState$ filtered to the vulnerability example, enabling precise cross-checking and self-correction.

\subsection{Query Optimization}
\label{s:design-opt}

We develop optimizations to mitigate query {underfitting} and {overfitting} and to improve query execution {efficiency}.

\subsubsection{Eliminating FPs for Precision}
\label{s:design-opt-fp}
When using the query $Q^t_i$ to analyze the project containing the input vulnerability example $e^t_i$, it might return false positives.
A false positive is often introduced at a certain code block $B_j$.
\sys thus similarly leverages the program state $PState$ to help the LLM reason about how the false positive is introduced and propose fixes.

In our evaluation, we take a best-effort manner to manually collect all true-positive vulnerabilities in a project as the example set $E^t$, therefore, all cases outside are considered false positives.
In practice, non-vulnerable code is much more prevalent than vulnerable code, so a report from an underdevelopment query is more like to be a false positive.
Recent research has shown it is possible to leverage an LLM to assess if a case is a false positive or not~\cite{iris}.
We leave this as future work.

\subsubsection{Generalization for Recall}
\label{s:design-opt-gen}

To mitigate the overfitting issue, \sys generalizes the queries to not just focus on a specific example.
This is especially important to make the queries capable of real-world detection later on.
As an initial step, we design a few heuristics to assess if a query is overfitting.
First, \sys checks whether the query relies on exact constant values (\eg{,} hardcoded variable names and string literals) that are unique to a specific example. 
Such reliance limits the applicability of the query to other contexts.
Second, \sys examines whether the query includes structural patterns or constraints that are overly specific to the layout of a particular program, such as deep AST chains or unique call sequences.
Once any such situation is found, \sys employs the LLM to abstract queries by asking it to
express the query in a more general form to resolve the identified overfitting situation.
Such generalization improves the reusability of queries and enables the detection of vulnerability variants.

\subsubsection{Merging for Efficiency}
\label{s:design-opt-merge}
After generating individual queries for different vulnerability examples, \sys performs query merging to consolidate detection queries and improve efficiency.
A naive approach would simply apply a logical \cc{OR} operation to combine all per-example queries---sequentially running each one independently.
However, this can lead to performance degradation, as many queries share redundant and repetitive conditions.
To address that, \sys introduces an LLM-assisted query merging stage.
All individual queries $Q^t_i$ are given to an LLM, which attempts to merge them into a single optimized query ($Q^t$) while preserving their detection capability.
The merged query is passed through the symbolic query validator for all vulnerability examples to verify its correctness and effectiveness (this process is omitted from \autoref{alg:query-gen}).
If the merged query fails validation on any of the examples, \sys further undergoes a similar feedback loop to iteratively improve the merging process.

\section{Implementation}
\label{s:impl}
We implemented our approach in a highly modular way and integrated it with Joern~\cite{joern} and CodeQL~\cite{codeql}.
The two leading query-based static analysis tools support a wide range of programming languages, including C/C++, C\#, Java, JavaScript, PHP, Python, Go, Ruby, Rust, Kotlin, Swift, and x86 binary.
The hardware requirements for running \sys align with the standard recommendations for Joern~\cite{joern-hw} and CodeQL~\cite{codeql-hw}, typically 16 GB RAM and 4 to 8 CPU cores.
We further show some important implementation details.

\PP{Instrumentation of Symbolic Validator}
We realized the runtime program state tracking by instrumenting the query execution runtime.
In particular, we enhanced the query parsing process to locate syntax errors.
For the other two types, we hooked the internal language interpreter.
Specifically, these query languages' interpreters typically contain a main interpretation loop that switches over the instruction types and invokes specific handlers.
We thus added additional hooks for selected handlers to collect variables and their runtime values.

\PP{Query Execution Server}
The iterative query generation would invoke the symbolic validator and static analysis runtime multiple times, and often operate on the same codebase, \ie{,} the project containing the input vulnerability example.
To improve the overall efficiency, we eliminate the repetitive project loading process by starting a local server to host the codebase for interactive, continuous query execution.

\section{Evaluation}
\label{s:eval}

We evaluate \sys to answer the following questions.
\squishlist
\item
RQ1: Can \sys generate valid queries?
\item 
RQ2: How effective are the \sys-generated queries for detecting vulnerabilities?
\item 
RQ3: How does \sys compare to expert-developed queries?
\item
RQ4: How does \sys compare to other SOTA LLM-based and specialized static analysis approaches?
\item %
RQ5: How effective is \sys in discovering new vulnerabilities in real-world applications?
\item 
RQ6: How does each component of \sys contribute to its performance?
\item
RQ7: What are the characteristics of the DSL subset used by \sys?
\squishend

\PP{Evaluated Vulnerability Types}
We evaluate \sys on 12 vulnerability types spanning four widely used languages (C/C++\footnote{We count C/C++ as one language since they share the same frontend in the underlying query-based static analysis tools.}, Java, PHP, and JavaScript).
While \sys supports 12 languages through Joern and CodeQL's infrastructure, we select these four for two reasons.
First, they are widely deployed in diverse security-critical domains, including systems software (C/C++), enterprise applications (Java), and web applications (PHP, JavaScript).
Second, they span memory-managed vs. unmanaged, statically-typed vs. dynamically-typed, and compiled vs. interpreted paradigms, enabling evaluation across diverse language characteristics.
The vulnerability types in them cover diverse attack vectors from memory safety issues to injection attacks (including use-after-free, buffer overflow, SQL injection, and XSS) and span both intraprocedural and interprocedural detection challenges.
Extending \sys to other languages or vulnerabilities requires no additional system modifications (\autoref{s:discussion}).

\PP{Dataset}
For each vulnerability type, we included known vulnerabilities for query generation and query evaluation.
Our dataset comprises \num{555} vulnerabilities across multiple projects, sourced from existing benchmarks (CWE-Bench-Java~\cite{iris}, PrimeVul~\cite{robin2025}, and Silent-Spring~\cite{silent-spring}) and public repositories (CVE database, GitHub issues, and patch commits).
To rigorously evaluate \sys, we partition our dataset into separate generation and testing sets to prevent data leakage.
Since vulnerabilities within the same project may share similarities, using them across query generation and evaluation could lead to data leakage.
We partition at the project level rather than at the vulnerability level to ensure complete isolation.
We randomly split projects into two sets following a roughly 40/60 ratio, with manual verification to ensure no overlap.
This results in \num{212} vulnerabilities in the generation set (\num{74} C/C++, \num{86} Java, \num{30} PHP, and \num{22} JavaScript) and \num{343} vulnerabilities in the testing set (\num{138} C/C++, \num{132} Java, \num{46} PHP, and \num{27} JavaScript) for effectiveness evaluation.

\PP{Tool Configuration}
\sys is a complex system with multiple configurable components, and we will systematically investigate their impact in \autoref{s:eval-ablation}.
Unless otherwise noted, \textit{our default setting} involves generating queries using the Claude 3.7 Sonnet model and Joern; the model is set with a temperature of 0.2 and a maximum token size of \num{10000}.
We ran the experiments on a CPU server equipped with dual 48-thread Intel Xeon processors and 187 GB RAM.

\begin{table}[t]
\centering
\caption{Query generation statistics. \cc{Covered}: the case is already detected by a previously generated query and no new query is generated; \cc{Failed}: threshold is reached without query convergence.}
\label{tab:generation}
\resizebox{\columnwidth}{!}{
\begin{tabular}{lrrrrrr}
\toprule
Vul. Type & Vul. & Success & Covered & Failed & Iteration & Time \\
\midrule
C/C++ UAF & 41 & 24 & 8 & 9 & 11.7 & 12.1h \\
C/C++ heap ov. & 33 & 19 & 6 & 8 & 10.9 & 9.4h \\
Java cmd. & 36 & 21 & 7 & 8 & 12.6 & 10.5h \\
Java SQLi & 21 & 12 & 4 & 5 & 9.2 & 6.0h \\
Java path trav. & 29 & 17 & 6 & 6 & 10.4 & 8.5h \\
PHP SQLi & 12 & 7 & 2 & 3 & 10.3 & 3.5h \\
PHP XSS & 8 & 5 & 1 & 2 & 11.8 & 2.5h \\
PHP type. & 4 & 2 & 1 & 1 & 13.5 & 1.0h \\
PHP deser. & 6 & 4 & 1 & 1 & 14.7 & 2.0h \\
JS proto. & 9 & 5 & 2 & 2 & 15.6 & 2.5h \\
JS cmd. & 4 & 2 & 1 & 1 & 10.8 & 1.0h \\
JS XSS & 9 & 5 & 2 & 2 & 12.4 & 2.5h \\
\midrule
Overall & 212 & 123 & 41 & 48 & 11.8 & 61.5h \\
\bottomrule
\end{tabular}
}
\end{table}

\subsection{RQ1: Query Generation}
\label{s:eval-querygen}
We first use \sys to generate detection queries from the vulnerabilities in the generation set.
For each vulnerability example, we use a maximum of 20 iterations (\cc{MaxN}), and \sys would abort if this threshold is reached without convergence.
We consider query generation successful if the resulting query can detect the input vulnerability example.
If a previously generated query already detects an example, we skip generating a new query for it to avoid redundancy.

\autoref{tab:generation} shows the detailed query generation statistics for each vulnerability type.
Out of 212 cases, \sys successfully handled 164 examples (77.4\%), either by generating new queries (123 cases) or by reusing previously generated queries that already covered them (41 cases).
The generated queries average 89.4 lines in length and demonstrate diverse static analysis capabilities.
Specifically, 112 queries incorporate data-flow or def-use tracking, 97 perform interprocedural analysis across function boundaries, and 54 employ control-flow conditions to filter spurious matches.
In addition, 83 queries leverage sanitization checks to reduce false positives.
This demonstrates that \sys generates sophisticated analysis logic beyond simple pattern matching.

The failed cases reached the iteration threshold without convergence.
These failures were primarily due to two reasons.
First, certain vulnerability patterns require intricate control-flow or data-flow specifications that could not be iteratively refined within the iteration threshold.
Second, limitations in the underlying static analysis tool's program representation, such as incomplete data flows or missing type information, prevented effective pattern generation.

\PP{Efficiency}
For successfully generated queries (excluding skipped cases), \sys required an average of 11.8 iterations per example.
Simpler patterns like Java SQLi converged quickly (avg. 9.2 iterations), while complex patterns like JavaScript prototype pollution required more iterations (avg. 15.6).
Overall, the majority of successful queries converged efficiently within the 20-iteration threshold.
Query generation required 61.5 hours total across all 123 successful examples.
Generation time varied by vulnerability type (1.0h to 12.1h), reflecting differences in both pattern complexity and the number of examples per type.
The generation is a one-time cost, and once generated, these queries can scan unlimited codebases instantly without additional time overhead.

\PP{API Cost}
Query generation under the default setting (Claude Sonnet 3.7) costs \$36 USD on average per vulnerability type for LLM inference.
Once generated, these queries can scan unlimited codebases without additional API cost.

\begin{table}[t]
\centering
\scriptsize
\caption{Vulnerability detection on the testing set.
\cc{Combined}: integrating both \sys-generated and expert-developed queries.
}
\label{tab:detection}
\resizebox{0.96\columnwidth}{!}
{
\begin{tabular}{lc|rr|rr|rr}
\toprule
& \multicolumn{1}{c}{} & \multicolumn{2}{c}{\sys} & \multicolumn{2}{c}{Experts} & \multicolumn{2}{c}{Combined} \\
\cmidrule(r){3-4} \cmidrule(lr){5-6} \cmidrule(l){7-8}
Vul. Type &  Total &  TP & FP &  TP & FP &  TP & FP \\
\midrule
C/C++ UAF   & 77  & 62  & 17  & 48  & 15  & 65  & 17  \\
C/C++ heap ov. & 61  & 46  & 9  & 44  & 14  & 50  & 15  \\
Java cmd. & 48 & 38  & 35  & 30  & 22  & 40  & 30  \\
Java SQLi  & 40  & 32  & 18  & 24  & 16  & 34  & 22  \\
Java path trav. & 44 & 35  & 28  & 24  & 20  & 37  & 26  \\
PHP SQLi  & 19 & 16 & 14 & 17 & 21 & 18 & 23 \\
PHP XSS   & 13 & 8  & 10 & 7  & 14 & 9  & 16 \\
PHP type. & 6  & 4  & 22 & 3  & 18 & 5  & 28 \\
PHP deser.& 8  & 5  & 4  & 4  & 6  & 7  & 7  \\
JS proto. & 11 & 10 & 8  & 7  & 12 & 11 & 13 \\
JS cmd.   & 5  & 5  & 17 & 5  & 15 & 5  & 20 \\
JS XSS    & 11 & 8  & 15 & 9  & 11 & 11 & 12 \\
\midrule
Overall & 343 & 269 & 197 & 222 & 184 & 292 & 229 \\
\bottomrule
\end{tabular}
}
\end{table}

\subsection{RQ2: Detection Effectiveness}
\label{s:eval-sys-detection}

We apply the generated queries to detect vulnerabilities in the testing set, and the per-type breakdown is shown as \cc{Total} in \autoref{tab:detection}.
Overall, out of 343 known vulnerabilities, \sys-generated queries successfully detected 269 cases with 197 false positives, achieving a detection rate of 78\%.

\PP{False Negatives}
We analyzed the 74 vulnerabilities (22\%) missed by \sys-generated queries. 
These false negatives stem from multiple root causes across different categories.
28 cases were caused by incomplete patterns, including missing taint source specifications (particularly involving customized third-party sources) or missing vulnerable operations, which could have been detected with more comprehensive patterns.
Another 15 cases involved incomplete data flows where the targets of dynamic function calls were missing from the project's program representations.
This occurred because the program representations failed to properly handle dynamic language features and indirect calls.
The remaining 31 cases were caused by inaccurate type information.
For instance, imprecise type information led to false negatives in PHP type juggling and JavaScript prototype pollution. 
These latter categories are limitations of the underlying static analysis infrastructure and cannot be resolved by refining the patterns alone.

\PP{False Positives}
\sys-generated queries produced reasonable false positives across the testing set, particularly for statically analyzing dynamic languages like PHP~\cite{tchecker} and JavaScript~\cite{silent-spring} and unmanaged languages like C/C++~\cite{sui2016svf}.
The false discovery rate (FDR), computed as $\frac{FP}{TP+FP}$, is 42.3\% for \sys.
As we will show in \autoref{s:eval-comparison}, expert-developed queries have a similar FDR of 45.4\%.

\PP{Analysis Time}  
\sys's queries took around 21.6 CPU hours in total to complete the analysis on all tested projects.
Details are presented in \autoref{tab:detection-efficiency} in \appautoref{s:appendix-table}.
The analysis is fully offline without LLM invocations.
Analysis time depends on application complexity and query design, but the overall duration remains practical for real-world use.

\subsection{RQ3: Comparison to Expert Queries} 
\label{s:eval-comparison}
Existing expert-developed queries for the same underlying tool represent an important baseline.
Since both \sys and expert queries operate on the same static analysis framework, this comparison removes the impact of tool-specific capabilities and design choices.
We evaluated the collected expert-developed queries on the same testing set.

\PP{Detection Performance}
Overall, \sys-generated queries achieved comparable performance to expert-developed queries, despite the latter requiring substantial security expertise and engineering effort.
As shown in \autoref{tab:detection}, expert-developed queries detected 222 true positives with 184 false positives, compared to \sys's 269 true positives with 197 false positives.
Our manual analysis reveals that \sys and expert-developed queries detected a substantial overlap in their detection capabilities, with both approaches identifying common vulnerabilities.
In several vulnerability types (C/C++ heap overflow, PHP deserialization, JavaScript command injection, JavaScript XSS), \sys-generated queries detected all vulnerabilities found by expert-developed queries.

\PP{Combination for Complementary Strengths}
However, our analysis of the queries reveals the different cognitive strengths of LLMs and human experts.
In particular, LLMs excel at capturing subtle low-level linguistic nuances that humans might overlook, while human experts are better at high-level conceptual reasoning and understanding broader vulnerability contexts.
For example, expert-developed queries \emph{all} overlooked PHP's case-insensitivity (\eg{,} \cc{echo} is the same as \cc{eCHo}), which was detected by \sys's strength in syntactic nuances. 
Conversely, expert queries showed superior design in identifying complex multi-step SQL injection~\cite{second-order} that require understanding broader application context.

This observation motivated us to investigate whether combining both sets of queries could lead to improved detection.
We evaluated the union of \sys-generated and expert-developed queries, resulting in a combined approach that detected 292 true positives with 229 false positives.
These results highlight that while \sys alone may not always outperform expert-developed queries, combining them yields a robust and highly effective detection strategy, achieving the highest detection coverage (85\% of all vulnerabilities).
Following this, we conducted a preliminary experiment to apply \sys to refine the expert-developed query for JavaScript prototype pollution as an example.
\sys ran for around one hour and automatically revised the query, which detected 3 previously missed vulnerabilities.
This highlights that \sys can not only automatically generate detection queries but also improve existing expert-developed queries.

\begin{table}[t]
\centering
\small
\caption{Overview of \numnewpattern new vulnerability patterns uncovered by \sys. The row marked with * summarizes 34 distinct \cc{phar}-based PHP deserialization operations.}
\label{tab:new-pattern}
\resizebox{\columnwidth}{!}
{
\begin{tabular}{llll}
\toprule
Vul. Type & Pattern & Functionality & Status \\
\midrule
General PHP  & \cc{htmlspecialchars} & Ctx-sensitive sanitization & Reported \\
PHP type. &  \cc{in_array}  & Implicit type casting & Merged \\
PHP type. &  \cc{array_search}  & Loose comparison & Merged \\
PHP type. &  \cc{array_flip}  & Implicit type casting & Reported\\
PHP type. &  \cc{case} in \cc{switch}  & Type coercion in case matching & Reported\\
*PHP deser.&  \cc{phar} & Object deserialization & Reported\\
PHP deser.&  \cc{copy} & Object deserialization & Reported\\
General JS   & \cc{Object.assign} & Data propagation & Reported\\
General JS   & \cc{Reflect.set} & Data manipulation & Merged\\
JS proto.    & \cc{Object.assign} & Data propagation & Merged \\
JS proto.    & \cc{Object.defineProperty} & New property manipulation & Reported\\
JS proto.    & \cc{obj.__defineGetter__} & New property manipulation & Merged \\
C/C++ UAF      & Reallocation via \cc{malloc} & Data propagation & Reported\\
\bottomrule
\end{tabular}
}
\end{table}

\PP{New Vulnerability Patterns}
Beyond comparable detection performance, \sys discovered \numnewpattern new vulnerability patterns that were missed by expert-developed queries, including both vulnerability-specific patterns and three general patterns applicable across entire programming languages, as shown in \autoref{tab:new-pattern}.
We define an atomic operation---such as a specific API call, expression type, or data flow---that is not recognized by expert-developed queries as a new pattern, and construct a minimal PoC vulnerability example to validate each new pattern.
For example, four new PHP type juggling mechanisms were found by \sys, including built-in functions \cc{array_search} and \cc{array_flip}, where implicit type casting could occur;
a new general way to propagate JavaScript data flows via \cc{Reflect.set}, which could benefit detection of different vulnerability types.
Although expert-developed queries for these fundamental vulnerability types had undergone multiple rounds of refinement, \sys still uncovered critical patterns that had been overlooked.
Each pattern represents a blind spot in the detection logic, and missing even one can result in an entire class of vulnerabilities going undetected.
This highlights the challenge of manually designing comprehensive query coverage, even for well-known bug types.
In contrast, \sys benefits from LLMs' ability to capture and enumerate subtle vulnerability behaviors.
We submitted the new patterns to the query maintainers, and to date, \numPR have been merged into upstream tools' repositories.

\PP{Development Effort}
We analyzed the Git commit history in the open-source repositories of the expert-developed queries, as shown in \autoref{tab:expert-time} in \appautoref{s:appendix-table}.
We find that experts often use multiple commits to develop and revise the queries. The commit history often spans from several days (\eg{,} 5 days for Java cmd.) to multiple weeks (\eg{,} 7 weeks for JS prototype pollution).
We acknowledge that Git commit history does not precisely reflect the actual, continuous engineering effort involved.
For instance, the time span between the first and last commit may overestimate or underestimate the true effort, as substantial work may have occurred before the first commit or between commits without being recorded.
Nevertheless, it serves as a reasonable proxy for comparing development timelines.
\sys, as an automated solution, has the advantage of running continuously and shortening the amount of time to construct queries.

\subsection{RQ4: Comparison to Other SOTA Approaches}  
\label{s:eval-sota}
We further assess \sys's effectiveness against the wider range of vulnerability detection methods by considering other SOTA approaches that employ different analysis frameworks and techniques.
Unlike \sys, which provides wide-ranging language and vulnerability type support, most prior solutions specialize in a single language or even a single vulnerability type.
We include a comprehensive set of both LLM-based and classic static analysis solutions.
We included four LLM-based baselines:
\squishlist
\item A pure LLM-based method that directly prompts an LLM with code. It is language- and vulnerability type-agnostic.
\item GRACE~\cite{lu2024grace} retrieves similar code examples from existing vulnerability benchmarks and uses them as demonstrations (few-show examples) in the prompt when evaluating new vulnerabilities. It is a stronger LLM-based baseline than the above approach. It supports C/C++.
\item IRIS~\cite{iris} uses LLMs to identify taint sources and sinks, which are then integrated into existing CodeQL queries for Java applications. 
Unlike \sys, which generates complete queries from scratch, IRIS only extends the taint specifications of existing queries.
\item KNighter~\cite{knighter} leverages LLMs to synthesize Clang Static Analyzer (CSA) based checkers for C/C++ programs.
\squishend
We select three representative specialized static analysis tools that leverage diverse techniques (\eg{,} LLVM-based pointer analysis, CFG-based dataflow analysis) and effectively represent the breadth of modern static analysis approaches:
\squishlist
\item LChecker~\cite{lchecker} performs pattern matching and type inference to identify PHP type juggling vulnerabilities.
\item Silent-Spring~\cite{silent-spring} detects JavaScript prototype pollution vulnerabilities.
\item Cred~\cite{yan2018spatio} analyzes LLVM bitcode to detect use-after-free vulnerabilities.
\squishend

\begin{table}[t]
\centering
\scriptsize
\caption{Comparison of \sys with other SOTA approaches. TPR = True Positive Rate; FDR = False Discovery Rate.}
\label{tab:sota-comparison}
\begin{tabular}{lrrrrrr}
\toprule
Dataset (Total Vuls.) & \\ %
~~Approach & TP &FN & FP &TPR & FDR\\
\midrule
\textbf{All Lang.} (343) & & & & & \\
~~Pure LLM & 185 & 158 & 182 & 53.9\% & 49.6\% \\
~~\sys & 269 & 74 & 197 & 78.4\% & 42.3\% \\
\midrule
\textbf{C/C++} (138)& & & & & \\
~~GRACE & 79 & 59 & 19 & 57.2\% & 19.4\% \\
~~KNighter & 57 & 81 & 16 & 41.3\% & 21.9\% \\
~~\sys & 108 & 30 & 26 & 78.3\% & 19.4\% \\
\midrule
\textbf{C/C++ UAF} (5) & & & & & \\
~~Cred & 4 & 1 & 4 & 80.0\% & 50.0\% \\
~~\sys & 4 & 1 & 6 & 80.0\% & 60.0\% \\
\midrule
\textbf{PHP type.} (6) & & & & & \\
~~LChecker & 6 & 0 & 37 & 100.0\% & 86.0\% \\
~~\sys & 4 & 2 & 22 & 66.7\% & 84.6\% \\
\midrule
\textbf{JS proto.} (11) & & & & & \\
~~Silent-Spring & 10 & 1 & 12 & 90.9\% & 54.5\% \\
~~\sys & 10 & 1 & 8 & 90.9\% & 44.4\% \\
\bottomrule
\end{tabular}
\end{table}

\subsubsection{Pure LLM-based Baseline}
This approach directly provides source code to an LLM and asks it to assess if there is any vulnerability.
Since LLM-based tools can hardly perform whole-project analysis for large codebases, we thus take the common practice by separately checking individual code snippets (source code file)~\cite{robin2025, aiware} with Claude 3.7 Sonnet.
To handle non-determinism across multiple LLM runs, we assign the final label based on majority agreement across three runs.

The evaluation results in \autoref{tab:sota-comparison} show significantly lower performance compared to \sys.
Out of 343 vulnerabilities in the testing set, pure LLM-based detection identified 185 true positives with 182 false positives.
This is primarily due to the lack of cross-file information, such as global variables and interprocedural function calls defined outside the current file, leading to many false positives and false negatives.
For instance, detecting JavaScript prototype pollution requires tracking data flow from external user input to object property manipulation, which is often infeasible when analyzing a single file in isolation.

\subsubsection{GRACE}
We evaluated GRACE~\cite{lu2024grace} on the 138 C/C++ vulnerabilities in our testing set.
GRACE enhances pure LLMs with few-shot examples.
GRACE detected 79 true positives with 19 false positives.
For comparison, the pure LLM baseline (without retrieval) on the same C/C++ vulnerabilities detected 73 true positives with 14 false positives.
\sys outperformed both approaches with a higher detection rate.
\sys particularly benefits from its reusable, interpretable queries that can be executed repeatedly without LLM API calls, whereas GRACE requires LLM inference for each new code snippet.

\subsubsection{KNighter}
We used KNighter~\cite{knighter} to generate CSA checkers from the 74 examples in our generation set.
As acknowledged in KNighter's paper, it focuses on intraprocedural analysis and does not support interprocedural analysis.
KNighter successfully generated valid analyzers for 28 cases.
We then applied these analyzers to our testing set, which found 57 true positives, substantially lower than \sys's 108 true positives.

The performance difference stems from multiple factors.
First, KNighter often generates overly specific checkers that rely on exact identifier names from training examples, which fail to generalize when testing and generation sets have different variable or function names.
Second, KNighter's reliance on the CSA, which operates on low-level program representations (\eg{,} memory regions) without providing sufficient high-level semantic abstractions (\eg{,} data flows, call graphs), limits its detection capability.
While the CSA has modest support for interprocedural analysis by design, expressing such analysis requires complex, error-prone specifications that are difficult for LLMs to synthesize correctly.
Moreover, KNighter relies solely on the LLM's pre-existing knowledge of the CSA API acquired during model pretraining, which may be incomplete or outdated.
This effectively restricts KNighter's analysis capability in detecting complex vulnerabilities like use-after-free.
In contrast, \sys explicitly provides the target DSL specification to the LLM, enabling it to learn and utilize high-level semantic abstractions (\eg{,} call graphs, data-flow edges, control dependencies) for interprocedural analysis.
Additionally, \sys's iterative refinement with generalization heuristics (\autoref{s:design-opt}) prevents overfitting to specific identifier names.
This design allows \sys to handle a broader range of vulnerability patterns across multiple languages and vulnerability types.

\begin{table}[t]
\centering
\footnotesize
\caption{Integrating \sys with IRIS on the Java dataset (132 positives). TPR = True Positive Rate; FDR = False Discovery Rate.}
\label{tab:sota-comparison-iris}
\resizebox{\columnwidth}{!}
{
\begin{tabular}{lrrrrr}
\toprule
Approach & TP &FN & FP &TPR & FDR\\
\midrule
CodeQL + CodeQL sources/sinks & 37 & 95 & 254 & 28.0\% & 87.3\% \\
CodeQL + IRIS sources/sinks & 68 & 64 & 324 & 51.5\% & 82.7\% \\
\sys + CodeQL sources/sinks & 55 & 77 & 256 & 41.7\% & 82.3\% \\
\sys + IRIS sources/sinks & 82 & 50 & 383 & 62.1\% & 82.4\% \\
\bottomrule
\end{tabular}
}
\end{table}

\subsubsection{IRIS}
\label{s:eval-sota-iris}
We evaluated IRIS~\cite{iris} on the 132 Java vulnerabilities from our testing set, more than half of which are sourced from the CWE-Bench-Java constructed in IRIS.
IRIS augments existing CodeQL queries by synthesizing project-specific taint sources and sinks using LLMs, while \sys generates complete queries from scratch.
To comprehensively assess their contributions, we evaluated different combinations of queries and taint specifications based on CodeQL.
Note that all configurations use CodeQL as the underlying analysis framework here (see \appautoref{s:eval-ablation-tool} for details on CodeQL query generation).
Specifically, we compared four configurations: (1) built-in CodeQL queries with CodeQL-defined sources/sinks (baseline), (2) built-in CodeQL queries with IRIS-synthesized sources/sinks (IRIS's approach), (3) \sys-generated queries with CodeQL-defined sources/sinks, and (4) \sys-generated queries with IRIS-synthesized sources/sinks (combined approach).
This setup allows us to separately measure the effectiveness of \sys's query generation capability and the complementary benefit of IRIS's taint specification synthesis.

The results are shown in \autoref{tab:sota-comparison-iris}.
\sys-generated queries consistently outperform CodeQL queries across both taint specification approaches.
Configuration (1) detected 37 true positives with 254 false positives, while configuration (2) improved to 68 true positives with 324 false positives, demonstrating IRIS's contribution to increasing recall.
Configuration (3) using \sys-generated queries with CodeQL sources/sinks detected 55 true positives with 256 false positives, already surpassing the CodeQL baseline.
The combined approach (4) achieved the best results with 82 true positives and 383 false positives, showing that \sys's query generation and IRIS's taint synthesis provide complementary benefits.
These results demonstrate that \sys generates more effective vulnerability detection queries, and can be further enhanced when combined with specialized taint specification techniques.

\subsubsection{LChecker}
LChecker reported 6 true positives and 37 false positives, while \sys reported 4 true positives and 22 false positives.
LChecker benefits from a more precise type inference algorithm than the built-in type information available in Joern, which helps improve recall.
However, in dynamic languages like PHP, many variable types cannot be resolved statically, leading LChecker to conservatively assign multiple possible types, resulting in over-approximation and more false positives.

\subsubsection{Silent-Spring}
We compare \sys and Silent-Spring on prototype pollution vulnerabilities.
Silent-Spring reported 10 true positives and 12 false positives, achieving the same recall as \sys but lower precision.
\sys achieves better precision by generating more targeted queries that reduce false positives.

\subsubsection{Cred}
Cred~\cite{yan2018spatio} relies on a compilation toolchain to analyze LLVM bitcode, while the vulnerabilities in the PrimeVul dataset are all individual functions and are not compilable or analyzable by Cred.
Thus, we can only compare Cred with \sys on the remaining UAF vulnerabilities in compilable, complete projects.
Cred reported the same number of four true positives as \sys, but with two fewer false positives.
This is because Cred leverages LLVM-based pointer analysis, which enables more accurate resolution of memory access patterns in C/C++ programs than Joern's graph-based analysis.
Nevertheless, \sys represents a significant step in automating pattern-based analysis.

\begin{table}[t]
\centering
\footnotesize
\caption{New vulnerabilities discovered by \sys.}
\label{tab:new}
{
\begin{tabular}{lcccc}
\toprule
Vul. Type & Total & Ack'ed & Fixed & Pending \\
\midrule
Java cmd. & 4 & 2 & 2 & 2\\
Java path trav. & 8 & 4 & 2 & 4 \\
PHP SQLi & 1 & 1 & 0 & 0\\
PHP XSS & 1 & 1 & 0 & 0\\
JS cmd. & 2 & 2 & 0 & 0 \\
JS Proto. & 9 & 4 & 1 & 5\\
\midrule
Total & 25 & 14 & 5 & 11\\
\bottomrule
\end{tabular}
}
\end{table}

\subsection{RQ5: New Vulnerabilities} 
\label{s:eval-new}
We applied the \sys-generated queries to discover new vulnerabilities in real-world applications.
\sys effectively discovered \numnewvul zero-day vulnerabilities spanning six vulnerability types across four languages, with a brief overview shown in \autoref{tab:new}.
We responsibly reported all discovered vulnerabilities to the respective project maintainers.
To date, 14 have been acknowledged as valid security issues and 5 have been fixed, with the remaining under review.
This demonstrates that \sys-generated queries are effective not only on benchmarks but also for discovering vulnerabilities in real-world applications.
Details on our project selection methodology, evaluation method, and a representative case study are presented in \appautoref{s:appendix-case-study}.

\begin{table}[t]
\centering
\caption{Query generation out of 212 examples. \cc{Covered}: the case is already detected by previously generated queries and no new query is needed;  \cc{Pass rate} = (Success + Covered) / Total.}
\label{tab:query-generation}
\begin{tabular}{lrrrr}
\toprule
Variant & Success & Covered & Failed & Pass Rate\\
\midrule
\sys & 123 & 41 & 48 & 77.4\%\\
DSL$_{100}$ & 89 & 17 & 106 &50.0\% \\
DSL$_{0}$ & 17 & 4 & 191 & 9.9\% \\
W/O FB$_{syntax}$ & 3 & 0 & 209 & 1.4\% \\
W/O FB$_{runtime}$ & 9 & 1 & 202 & 4.7\% \\
W/O FB$_{semantic}$ & 10 & 1 & 201 & 5.2\% \\
\bottomrule
\end{tabular}
\end{table}

\subsection{RQ6: Ablation Study}   
\label{s:eval-ablation}
In our ablation study, we assess how key components in \sys affect both query generation, subsequent vulnerability detection performance, and the impact of query merging.
We also evaluate different LLM models and other underlying static tools in \appautoref{s:appendix-ablation}.

\subsubsection{Query Generation}
\label{s:eval-ablation-querygen}
We assess how DSL and feedback mechanisms contribute to successful query generation.

\PP{DSL}
The DSL specification guides the LLM in generating valid queries using appropriate API constructs.
We compare \sys with DSL subsetting against two variants: (1) DSL$_{100}$ with a complete DSL specification (no subsetting) and (2) DSL$_{0}$ with no specification.
We use the same iteration threshold as in \autoref{s:eval-querygen} and calculate how many input examples in the generation set lead to successful queries.
As shown in \autoref{tab:query-generation}, DSL$_{0}$ achieved only 9.9\% pass rate with 17 successful queries, confirming that DSL guidance is essential.
DSL$_{100}$ achieved 50\% query generation success under the same iteration threshold.
The high generation success rate of 77.4\% in \sys highlights the importance of DSL and subsetting.
We will characterize the DSLs in terms of what is preserved or removed in \autoref{s:eval-dsl}.

\PP{Feedback Mechanism}
The feedback mechanism addresses three types of errors sequentially.
We thus separately disable each feedback type: (1) W/O FB$_{syntax}$ removes syntax error feedback, (2) W/O FB$_{runtime}$ removes runtime error feedback, and (3) W/O FB$_{semantic}$ removes semantic validation feedback.
Due to this dependency structure, a query must resolve syntax and runtime errors before semantic validation becomes meaningful.
Without syntax feedback, the LLM continues to generate syntactically invalid queries across iterations, often hallucinating non-existent APIs or malformed constructs, achieving a 1.4\% ultimate pass rate.
Disabling runtime feedback allows the LLM to produce syntactically valid but non-executable queries that repeatedly fail with runtime errors, reaching a 4.7\% pass rate.
Disabling semantic feedback permits queries to execute successfully but misses the intended vulnerabilities, as the LLM lacks guidance on whether the query logic correctly captures the vulnerability pattern, achieving a 5.2\% pass rate.
In contrast, \sys with all three feedback mechanisms achieves 77.4\% pass rate, demonstrating their critical role in guiding iterative refinement toward valid and effective queries.

\begin{table}[t]
\centering
\caption{Vulnerability detection performance (F1 scores).}
\label{tab:ablation}
\begin{tabular}{lrrrr}
\toprule
Vul. Type & \sys & DSL$_{100}$ & W/O Gen. & W/O FP \\
\midrule
C/C++ UAF   & 0.53 & 0.28 & 0.35 & 0.31 \\
C/C++ heap ov.& 0.46 & 0.23 & 0.22 & 0.27 \\
Java cmd.   & 0.50 & 0.36 & 0.33 & 0.29 \\
Java SQLi   & 0.55 & 0.30 & 0.29 & 0.26 \\
Java path traversal & 0.50 & 0.22 & 0.28 & 0.25 \\
PHP SQLi    & 0.65 & 0.38 & 0.37 & 0.54 \\
PHP XSS     & 0.52 & 0.23 & 0.25 & 0.32 \\
PHP type.   & 0.25 & 0.15 & 0.23 & 0.19 \\
PHP deser.  & 0.59 & 0.41 & 0.28 & 0.34 \\
JS proto.   & 0.69 & 0.54 & 0.28 & 0.33 \\
JS cmd.     & 0.37 & 0.29 & 0.32 & 0.18 \\
JS XSS      & 0.48 & 0.33 & 0.32 & 0.23 \\
\midrule
Average     & 0.51 & 0.31 & 0.29 & 0.29 \\
\bottomrule
\end{tabular}
\end{table}

\subsubsection{Vulnerability Detection Performance}
\label{s:eval-ablation-detection}
We evaluate how different components affect detection effectiveness on the subset of variants that successfully generate queries.
We focus on three key variants of \sys: (1) DSL$_{100}$ uses the complete DSL specification instead of our curated subset, (2) W/O Gen. disables the generalization heuristics that detect and fix overfitting to specific examples, and (3) W/O FP disables the false positive elimination feedback that refines queries to improve precision.
We exclude DSL$_{0}$ and the three feedback ablations (W/O FB$_{syntax}$, W/O FB$_{runtime}$, W/O FB$_{semantic}$) from this analysis, as their low query generation rates (below 10\%) result in insufficient queries for meaningful detection performance comparison.
We measure detection effectiveness using F1 score ($2\cdot\frac{Prec.\cdot Rec.}{Prec.+Rec.}$), which balances precision and recall and is well-suited for imbalanced datasets where non-vulnerable code vastly outnumbers vulnerable code.

As shown in \autoref{tab:ablation}, \sys achieves an average F1 score of 0.51 across all vulnerability types, substantially outperforming all ablation variants.
DSL$_{100}$ achieves a 0.31 F1 score, demonstrating that the complete DSL specification hinders detection despite reasonable query generation rates (49.1\%).
Both W/O Gen. and W/O FP achieve a 0.29 F1 score, confirming that generalization heuristics and false positive elimination are equally critical for effective detection.
Without generalization, queries overfit to specific code structures in the input examples, failing to detect structural variations of the same vulnerability pattern.
Without false positive elimination, queries introduce substantially more spurious warnings when analyzing test projects, degrading precision despite maintaining recall.
These results confirm that each component of \sys (DSL subsetting, feedback-driven refinement, generalization heuristics, and false positive elimination) contributes substantially to both successful query generation and effective vulnerability detection.

\subsubsection{Query Merging and Efficiency}
\label{s:eval-ablation-merge}
The query merging stage in \sys improves analysis efficiency by consolidating multiple per-example queries into a single unified query.
We designed a baseline that sequentially runs each per-example query without merging and measured the cumulative query execution time on the testing set.
The analysis time for each vulnerability type is presented in \autoref{tab:detection-efficiency} in \appautoref{s:appendix-table}.
Without merging, the analysis required over 130 hours (with four vulnerability types reaching the 24-hour timeout), compared to only 21.6 hours with merging.
This 6\x improvement demonstrates the necessity of query merging for practical vulnerability detection at scale.

\subsection{RQ7: DSL Subset Characteristics}
\label{s:eval-dsl}
We evaluate the characteristics of the DSL subset extracted by \sys and its impact on query generation effectiveness.
Joern exposes a total of \num{4591} DSL APIs, while CodeQL exposes \num{12934} DSL APIs.
The subsetting process reduces Joern's DSL from \num{4591} to \num{2387} APIs (48\% reduction) and CodeQL's from \num{12934} to \num{4268} APIs (67\% reduction), substantially reducing the search space while maintaining sufficient expressiveness.

\PP{Removed APIs}
We manually analyzed 50 randomly sampled removed APIs to understand why they could be safely excluded.
The majority (45 out of 50) of removed features had equivalent alternatives within the retained subset.
For example, Joern's \cc{whereNot()} can be expressed as \cc{where(not)}.
This redundancy arises because DSL features are typically introduced to simplify usage and improve expressiveness, yet some offer functionality that can be composed from other constructs.
Two removed features were non-essential debugging or printing operations unnecessary for production queries.
The remaining three were language-specific operations irrelevant to our target languages (\eg{}, \cc{address} used only for binary analysis).
All 50 sampled removals were confirmed as truly redundant, demonstrating the reliability of LLMs in identifying non-essential DSL features.

\PP{Retained APIs}
To complement the analysis of removed APIs, we also analyzed 50 randomly sampled retained APIs.
We found that 88\% (44 out of 50) were necessary for expressing the vulnerability patterns in our dataset, such as referring to operation types and fetching functions.
The remaining 6 APIs, while potentially useful, were not strictly necessary given the available alternatives.
Nevertheless, conservatively retaining extra features is preferable to missing essential ones.

\section{Discussion}
\label{s:discussion}

\PP{Extensibility}
While we evaluated four languages (C/C++, Java, PHP, JavaScript), our implementation supports 12 languages through Joern and CodeQL, including Python, C\#, Ruby, and Go.
We selected the four languages based on their prevalence in security-critical applications and the availability of diverse vulnerability benchmarks.
We expect \sys to generalize to the remaining languages given their shared DSL infrastructure and query generation pipeline, though language-specific features (\eg{}, Python's dynamic typing, C\#'s reflection) may require future validation.
Similarly, while we evaluated 12 vulnerability types, \sys can generate patterns for new types given only a small number of examples, without requiring design changes.

Beyond Joern~\cite{joern, phpjoern} and CodeQL~\cite{codeql}, \sys's core workflow of DSL extraction, iterative refinement, and symbolic validation can extend to other query-based frameworks such as Amazon CodeGuru~\cite{codeguru} and Datalog-based systems~\cite{scholz2016fast}.
For non-query-based approaches using different pattern representations (\eg{}, abstract interpretation rules), adapting \sys would require specifying how patterns are expressed and validated in those frameworks.

\PP{Limitations}
\sys has a few limitations that warrant future research.
First, \sys's effectiveness relies on the underlying tool's capabilities since \sys generates expressive patterns executed by the tools.
If a tool lacks essential program representations or capability (\eg{}, fine-grained pointer analysis), \sys cannot guarantee to generate effective and precise vulnerability queries.
Second, \sys-generated queries produce false positives, similar to expert-developed queries.
A possible way to mitigate this is to leverage LLMs to validate the specific code snippets flagged in the results, as demonstrated in early work~\cite{llmxcpg}.
We currently do not include this component as we believe it is orthogonal to the main focus of generating vulnerability patterns.
Third, \sys currently follows a fixed flow that generates and then executes queries.
Some sophisticated vulnerabilities may require more dynamic composition or interplay between LLM and static analysis for complex real-world scenarios.
For example, leveraging LLMs for analyzing intermediate results and dynamically adjusting query strategies based on partial findings.
Future work could combine multiple analysis frameworks to leverage their complementary strengths and develop techniques to reduce false positives while maintaining recall.

\section{Related Work}
\label{s:relwk}

\PP{Learning-based Detection}
Transformer-based models have been widely applied to vulnerability detection, including encoder-only~\cite{Feng2020CodeBERTAP}, encoder-decoder~\cite{Wang2021CodeT5IU, Wang2023CodeT5OC}, and decoder-only~\cite{codellama} architectures.
To improve performance, researchers have explored various fine-tuning strategies, such as domain-specific pretraining~\cite{Ding2023Concord}, instruct-tuning~\cite{Mao2024TowardsEV, Chang2024Instruct}, and supervised fine-tuning~\cite{Zhu2024Finetune, WANG2024103994}.
However, recent studies~\cite{Ullah2023LLMsCR} show that they require a large dataset for training and often present reduced effectiveness on under-represented vulnerabilities and large codebases.
Compared to these approaches, \sys benefits from symbolic reasoning and achieves superior scalability.

\PP{Neuro-symbolic Program Analysis}
Neuro-symbolic program analysis integrates neural and symbolic components.
IRIS~\cite{iris} uses LLMs to infer project-specific taint specifications for existing Java queries and is complementary to \sys that generate queries from scratch.
KNighter~\cite{knighter} generates Clang static analyzers to find recurring vulnerabilities within the scope of a function, and its current design does not support interprocedural analysis. 
Other works use neural components to assist symbolic validation, such as WAP~\cite{wap}, or decompose analysis into smaller property checks as in LLMSA~\cite{llmsa}.
LLift~\cite{llift} applies LLMs to reason about post-conditions in use-before-initialization bugs in the Linux kernel.
LLMxCPG~\cite{llmxcpg} constructs program slices and then leverages an LLM to validate.
\sys differs by aiming to automatically generate vulnerability queries from scratch, and can complement existing approaches by reducing manual engineering effort.

\PP{General Static Analysis}
Static vulnerability detection includes a broad range of techniques beyond pattern-based approaches.
Methods such as symbolic execution~\cite{baldoni2018survey, dudina2017using}, model checking~\cite{yang2006using, su2021model}, and formal verification~\cite{hasan2015formal} analyze program behavior against formal specifications or mathematical models, aiming to logically prove security properties or identify bugs.
Other general techniques, such as dataflow analysis frameworks based on lattices~\cite{cousot1977abstract}, apply abstract interpretation to systematically reason about program behavior and detect potential vulnerabilities.
Unlike pattern-based tools, these approaches do not depend on manually crafted vulnerability signatures.
However, despite their precision and theoretical rigor, formal and dataflow-based methods often face significant scalability challenges when applied to large or complex codebases.

\section{Conclusion}
\label{s:conclusion}
This work presents \sys, a novel neuro-symbolic approach that combines the complementary strengths of LLMs and static analysis to enable scalable, automated vulnerability detection.
By leveraging DSL subsetting and trace-driven symbolic validation, \sys automatically generates vulnerability detection patterns from examples.
\sys-generated queries achieved comparable detection performance to state-of-the-art approaches while reducing engineering effort from weeks to hours.
Beyond comparable performance, \sys discovered \numnewpattern new vulnerability patterns missed by experts and detected \numnewvul previously unknown vulnerabilities in real-world applications.
By lowering the barrier to developing custom vulnerability detectors, we believe this neuro-symbolic approach represents a promising direction to make security analysis more accessible and adaptable

\section*{Ethics considerations}
We identify two primary ethical considerations in this work. First, the new vulnerabilities could potentially have impacts on the software, its developers, and users.
To mitigate the impacts, all newly discovered vulnerabilities were responsibly disclosed to the respective developers and maintainers following coordinated disclosure practices before submission.
We did not make the disclosure public to avoid potential harm to its users.
Second, using LLMs in security-critical contexts carries inherent risks.
In our setting, all LLM-generated queries were validated through rigorous testing and human review before deployment.
We also notified the upstream to work together to improve the queries.
Our evaluation uses only publicly available vulnerability benchmarks and open-source projects.
No personal identifiable information or sensitive data was collected or processed in this research.

\section*{LLM usage considerations}
In this work, LLMs were used for editorial purposes in this manuscript, and all outputs were inspected by the authors to ensure accuracy and originality. In particular, we independently conducted our literature review to identify the bottlenecks of pattern-based static analysis, developed the \sys system, designed and executed all experiments, and wrote the initial draft of the paper. After the initial draft was completed, we used LLMs to polish part of the text for clarity, improve grammatical correctness, and enhance readability. We ensure that all technical content, methodological decisions, experimental results, analysis, and conclusions remain entirely the original work of the authors.

Additionally, \sys leverages LLMs (primarily Claude 3.7 Sonnet, with comparative experiments using GPT-4o) to automatically generate vulnerability detection queries from examples. All LLM-generated queries were rigorously validated through our symbolic query validator and tested against ground-truth vulnerability datasets. Our use of commercial, closed-source models may introduce reproducibility challenges. First, LLM outputs exhibit inherent non-determinism. Second, commercial API providers may update or deprecate specific model versions without notice, preventing future researchers from accessing the exact Claude 3.7 Sonnet version used in our experiments. To mitigate these limitations, we document exact model versions and parameters (temperature=0.2, max\_tokens=10,000), provide simplified prompts in Appendix D, and demonstrate the applicability of \sys across different LLMs. We also run the query generation pipeline multiple times to ensure result stability, and we will release all generated queries in our artifact for independent verification.

We do not train any models in this work. \sys only uses a small number of vulnerability examples to generate detection patterns through in-context learning. These examples are sourced from existing open-source benchmarks (CWE-Bench-Java, PrimeVul, and Silent-Spring) and public repositories (CVE database, GitHub issues, and patch commits), all of which are publicly available and do not raise concerns regarding consent, data holder rights, or intellectual property. Since we rely on pre-trained commercial models, our environmental footprint is limited to inference costs totaling 61.5 hours of LLM API usage, as discussed in Section 5.1. We minimized our footprint by skipping cases already covered by existing generated queries (41 out of 212 cases were skipped for this reason). This cost is justified by our research goal of automating vulnerability pattern generation that previously required weeks of manual expert effort. Once queries are generated, they can be reused for future vulnerability detection without additional inference cost or recurring environmental impact.

\bibliographystyle{ACM-Reference-Format}
\bibliography{p,conf}
\appendix

\section{Details of New Vulnerability Detection}
\label{s:appendix-case-study}

\subsection{Methodology}
We searched GitHub for real-world open-source projects using keywords (\eg{,} \cc{web}, \cc{data processing}, \cc{utilities}) filtered by programming language.
We selected actively maintained projects (with commits in the last year) that were not in our training or testing datasets.
In total, we analyzed 42 projects across the four languages.
The details of the project list are provided in our artifact.
We automatically scanned each project's latest version with \sys-generated queries, manually triaged all findings, and responsibly disclosed confirmed vulnerabilities to maintainers.

\subsection{Case Study}
\autoref{code:example} shows a representative new prototype pollution vulnerability discovered by \sys in \cc{es2015-proxy}~\cite{es2015-proxy}.
The \cc{mergeOptions} function iterates over the properties of the \cc{source} object and assigns the property values to the \cc{target} object.
When \cc{source} is crafted to include the special property \cc{"\_\_proto\_\_"}, the \cc{mergeOptions} function inadvertently assigns its value to overwrite \cc{target["\_\_proto\_\_"]}, which refers to the object's prototype.
Specifically, the proof-of-concept (PoC) payload we created modifies the implementation of the \cc{toString} method of \cc{target["\_\_proto\_\_"]}.
This modification impacts not only \cc{target} but also all objects that inherit from the foundational object prototype \cc{Object.prototype}, such as the \cc{newObj} on line 13.
Detecting this vulnerability requires capturing the \cc{Object.assign} in the pattern, which was missed by expert-developed queries.
\begin{listing}[h]
    \begin{Verbatim}[commandchars=\\\{\},codes={\catcode`\$=3\catcode`\^=7\catcode`\_=8\relax}]
\PY{k+kd}{function}\PY{+w}{ }\PY{n+nx}{mergeOptions}\PY{p}{(}\PY{n+nx}{target}\PY{p}{,}\PY{+w}{ }\PY{n+nx}{source}\PY{p}{)}\PY{+w}{ }\PY{p}{\PYZob{}}
\PY{+w}{  }\PY{k}{for}\PY{+w}{ }\PY{p}{(}\PY{k+kd}{let}\PY{+w}{ }\PY{n+nx}{key}\PY{+w}{ }\PY{o+ow}{in}\PY{+w}{ }\PY{n+nx}{source}\PY{p}{)}\PY{+w}{ }\PY{p}{\PYZob{}}
\PY{+w}{    }\PY{n+nb}{Object}\PY{p}{.}\PY{n+nx}{assign}\PY{p}{(}\PY{n+nx}{target}\PY{p}{[}\PY{n+nx}{key}\PY{p}{]}\PY{p}{,}\PY{+w}{ }\PY{n+nx}{source}\PY{p}{[}\PY{n+nx}{key}\PY{p}{]}\PY{p}{)}\PY{p}{;}
\PY{+w}{  }\PY{p}{\PYZcb{}}
\PY{+w}{  }\PY{k}{return}\PY{+w}{ }\PY{n+nx}{target}\PY{p}{;}
\PY{p}{\PYZcb{}}
\PY{c+c1}{// Proof\PYZhy{}of\PYZhy{}Concept (PoC)}
\PY{k+kd}{const}\PY{+w}{ }\PY{n+nx}{obj}\PY{+w}{ }\PY{o}{=}\PY{+w}{ }\PY{p}{\PYZob{}}\PY{p}{\PYZcb{}}
\PY{k+kd}{const}\PY{+w}{ }\PY{n+nx}{payload}\PY{+w}{ }\PY{o}{=}\PY{+w}{ }\PY{l+s+s1}{\PYZsq{}\PYZob{}\PYZdq{}\PYZus{}\PYZus{}proto\PYZus{}\PYZus{}\PYZdq{}: \PYZob{}\PYZdq{}toString\PYZdq{}: \PYZdq{}Hacked\PYZdq{}\PYZcb{}\PYZcb{}\PYZsq{}}\PY{p}{;}
\PY{n+nx}{mergeOptions}\PY{p}{(}\PY{n+nx}{obj}\PY{p}{,}\PY{+w}{ }\PY{n+nb}{JSON}\PY{p}{.}\PY{n+nx}{parse}\PY{p}{(}\PY{n+nx}{payload}\PY{p}{)}\PY{p}{)}\PY{p}{;}

\PY{k+kd}{const}\PY{+w}{ }\PY{n+nx}{newObj}\PY{+w}{ }\PY{o}{=}\PY{+w}{ }\PY{p}{\PYZob{}}\PY{p}{\PYZcb{}}\PY{p}{;}
\PY{n+nx}{console}\PY{p}{.}\PY{n+nx}{log}\PY{p}{(}\PY{n+nx}{newObj}\PY{p}{.}\PY{n+nx}{toString}\PY{p}{)}\PY{p}{;}\PY{+w}{ }\PY{c+c1}{// \PYZdq{}Hacked\PYZdq{}}
\end{Verbatim}

    \caption{A JavaScript prototype pollution vulnerability.}
    \label{code:example}
\end{listing}

\section{Additional Ablation Study}
\label{s:appendix-ablation}

\subsection{Using GPT-4o}
\label{s:eval-ablation-model}
\sys is not limited to Claude 3.7 Sonnet.
We further experimented with GPT-4o~\cite{gpt4o} to assess \sys's applicability to other LLMs.
We evaluated GPT-4o on the same generation and testing sets using Joern as the static analysis backend.
In our experiments, GPT-4o successfully generated queries for 67 out of 212 vulnerability examples in the generation set, compared to 123 for Claude 3.7 Sonnet.
On the testing set, GPT-4o achieved an average F1 score of 0.37, while \sys with Claude 3.7 Sonnet achieved 0.51.
The performance gap was particularly noticeable on complex vulnerability types requiring multi-step analysis, such as prototype pollution.
Our experience suggests that Claude 3.7 Sonnet has better capabilities for code generation tasks than GPT-4o, particularly in understanding and generating queries in domain-specific languages.
While we did not integrate additional models due to cost constraints, we believe \sys's approach is generally applicable to other advanced LLMs.

\subsection{CodeQL Query Generation} 
\label{s:eval-ablation-tool}
Our evaluation uses Joern as the default static analysis tool. 
We further evaluated \sys with CodeQL to demonstrate its adaptability across different static analysis platforms, also using Claude 3.7 Sonnet.
Since CodeQL does not support PHP analysis, we evaluated it on the remaining 8 vulnerability types (182 examples) across C/C++, Java, and JavaScript.
\sys with CodeQL successfully handled 97 out of 182 examples (53.3\%), including generating new queries (82 cases) and reusing previously generated queries (15 cases).
In comparison, \sys with Joern on these same three languages handled 141 out of 182 examples (77.5\%), including 105 new queries and 36 reusing cases.
CodeQL's richer but more complex DSL resulted in lower query generation success compared to Joern.

We then applied the generated queries to the testing set with 297 vulnerabilities across the three languages.
CodeQL-based queries detected 172 true positives with 319 false positives, while Joern-based queries detected 236 true positives with 147 false positives.
The higher false positive rate for CodeQL stems partially from lower query generation success, which limited opportunities for false positive refinement during the iterative generation process.
Note that the Java queries generated with CodeQL are also used in the IRIS comparison (\autoref{s:eval-sota-iris}).

\section{Additional Result Tables}
\label{s:appendix-table}
In this section, we present additional result tables mentioned in the main text.
Specifically,
\autoref{tab:detection-efficiency} shows the efficiency comparison with or without query merging.
\autoref{tab:expert-time} shows the development time of expert-developed queries.

\begin{table}[th]
\centering
\scriptsize
\caption{Analysis time (in hours) for scanning the testing set with and without the merging component. 24+ indicates a timeout after 24 hours. Total time: 21.6 hours with merging vs. >130 hours without merging (4 timeouts).}
\label{tab:detection-efficiency}
\resizebox{\columnwidth}{!}
{
\begin{tabular}{l|ccccccc}
\toprule
Vul. Type & C/C++ UAF & C/C++ heap ov. & Java cmd. & Java SQLi \\
\midrule
W/ Merging & 1.3 & 1.4 & 1.8 & 1.7\\
W/O Merging & 24+ & 24+ & 24+ & 24+ \\
\midrule
Vul. Type & Java path trav. & PHP SQLi & PHP XSS & PHP type. \\
\midrule
W/ Merging & 3.8 & 2.2 & 4.9 & 1.2\\
W/O Merging & 18.4 & 5.6 & 7.5 & 6.2 \\
\midrule
Vul. Type & PHP deser. & JS proto. & JS cmd. & JS XSS \\
\midrule
W/ Merging & 0.5 & 1.3 & 0.4 & 1.1 \\
W/O Merging & 5.4 & 2.8 & 3.4 & 2.7 \\
\bottomrule
\end{tabular}
}
\end{table}

\begin{table}[th]
\centering
\scriptsize
\caption{Expert query development time from Git commit history. \cc{NA} denotes that the time is not available.}
\label{tab:expert-time}
{
\begin{tabular}{l|cccccccc}
\toprule
Vul. Type & C/C++ UAF & C/C++ heap ov. & Java cmd. & Java SQLi \\
\midrule
Dev. Time & 6w & 4w & 5d & 3w \\
\midrule
Vul. Type & Java path & PHP SQLi & PHP XSS & PHP type. \\
\midrule
Dev. Time & 2w & 2d & 3d & 7d \\
\midrule
Vul. Type & PHP deser. & JS proto. & JS cmd. & JS XSS \\
\midrule
Dev. Time & 5w & 7w & NA & NA \\
\bottomrule
\end{tabular}
}
\end{table}

\section{Prompts Used in \sys}
\label{s:appendix-prompts}
In this section, we show the prompts used in \sys.

\promptbox{DSL Grammar Extraction.}{
    You are an expert on \pt{tool-name}. Your task is to analyze the documentation below and extract the grammar of the domain-specific query language. \\
    \\
    1. Identify the core syntactic elements (keywords, operators, predicates). \\
    2. Determine the composition rules and how elements combine. \\
    3. Formalize the grammar in Backus-Naur Form (BNF). \\
    \\
    Please provide your analysis step-by-step, then present the final BNF grammar. \\
    \\
    Documentation: \{\pt{documentation}\}
}

\promptbox{DSL Semantics Extraction.}{
    You are an expert on \pt{tool-name}. Your task is to analyze the source code implementations and document the functionality of each domain-specific API. \\
    \\
    For each API listed below: \\
    1. Identify its purpose and core functionality \\
    2. Describe its input parameters and return values \\
    3. Explain its behavior and any side effects \\
    4. Note any important constraints or edge cases \\
    \\
    Please provide a structured summary for each API. \\
    \\
    APIs to analyze: \{\pt{API list}\} \\
    Source code: \{\pt{source code}\}
}

\promptbox{DSL Subsetting.}{
    You are an expert on \pt{tool-name}. \\
    The complete DSL for \pt{tool-name} is large and complex, containing many features that may not be necessary for vulnerability detection. \\
    Your task is to extract a minimal but sufficiently expressive subset of the DSL for detecting vulnerabilities in the programming language \pt{L}. \\
    Full DSL specification: \{\pt{complete DSL}\} \\
    Please follow this process: \\
    1. Analyze the essential operations needed for vulnerability detection (e.g., tracking data flow, identifying sources/sinks, checking conditions) \\
    2. Review the sample vulnerabilities and identify which DSL features are required to express their detection patterns \\
    3. Categorize DSL features into: (a) essential for vulnerability detection, (b) useful but not critical, (c) unnecessary for this use case \\
    4. Select the minimal subset that maintains sufficient expressiveness \\
    5. Justify your selection by explaining how the subset covers the key detection patterns \\
    \\
    Provide your analysis for each step, then present the final DSL subset with a rationale. \\
    You can also refer to the few existing query examples.
}

\clearpage

\promptbox{Query Generation.}{
    You are an expert in \pt{tool-name}. Your task is to write a vulnerability detection query for \pt{vulnerability-type}. \\
    Available DSL: \{\pt{DSL syntax and semantics}\} \\
    \\
    Please follow this step-by-step process: \\
    1. Analyze the vulnerability pattern: Identify sources (user input), sinks (dangerous functions), and any sanitization gaps \\
    2. Break down the detection into subtasks (e.g., find sources, track data flow, identify sinks, check for sanitization) \\
    3. For each subtask, write a corresponding DSL query fragment with a brief explanation \\
    4. Combine all fragments into a complete, cohesive query \\
    5. Verify the query logic covers the vulnerability pattern \\
    Reference vulnerability example: \{\pt{vulnerability example}\} \\
    Example queries for similar vulnerabilities: \{\pt{query examples}\} \\
    \\
    Think through each step, then provide your final query.
}

\promptbox{Query Refinement.}{
    You are an expert on \pt{tool-name}. Your previous query for \pt{vulnerability-type} encountered errors. \\
    Previous query: \{\pt{previous query}\} \\
    Error type: \pt{syntax errors} / \pt{execution exceptions} / \pt{semantic errors} \\
    Error details: \{\pt{error messages}\} \\
    Execution state: \{\pt{execution state}\} \\
    \\
    Please analyze the error: \\
    1. Identify the root cause of the error \\
    2. Explain what went wrong in your previous approach \\
    3. Propose a fix with justification \\
    4. Generate the corrected query \\
    Provide your reasoning, then the revised query.
}

\end{document}